\documentclass[12pt,showpacs,preprintnumbers,amsmath,aps,amssymb,epsfig]{revtex4}
\usepackage{graphicx}% Include figure files
\usepackage{dcolumn}% Align table columns on decimal point
\usepackage{bm}% bold math
\usepackage{color}

\begin{document}

\title{Anti-self-dual gravity and supergravity from a pure connection formulation.}
\author{J.E. Rosales-Quintero}
\email{erosales@fisica.ugto.mx} \affiliation{Departamento de
F\'isica, DCI, Campus Le\'on, Universidad   de Guanajuato, A.P.
E-143, C.P. 37150, Le\'on, Guanajuato, M\'exico.}

\begin{abstract}

We introduce a  complex pure connection action  with constraints
which is diffeomorphism and gauge invariant. Taking as an internal
group $SU(2)$, we obtain, from the equations of motion,
anti-self-dual Einstein spaces  together with the zero torsion
condition thanks to Bianchi identity.  By applying the same
procedure, we take as internal symmetry the super group $OSp(1|2)$
and by means of the Bianchi identity and integrability conditions,
the equations of motion are those that come from anti-self-dual
supergravity $N=1$ with cosmological constant sector.

\end{abstract}
\date{\today}
\pacs{00} \keywords{}
%\preprint{}
%\maketitle

\maketitle

\section{ INTRODUCTION}
It was Einstein's great achieve that gravity is a manifestation of
spacetime curvature. In Riemannian geometry, a spacetime with
curvature is describe by the metric and Einstein therefore used it
to describe gravity. Being the metric the fundamental variable
which determines spacetime intervals, and provides the causal
structure to which all interactions including gravity  itself must
conform, it is astonishing that gravity can be reformulated in
such a way that the metric doesn't play  the central role, instead
it becomes a derived object, and then, as  we do not
need a pre-existing metric,  the theory becomes background  free.\\
In middle of 1970, Pleba\'nski\cite{Plebanski} adopted the
viewpoint that the two forms are basic variables, and he exhibited
a first order Palatini type action for complex General Relativity
(GR) in which the field variables are certain  triple of
self-dual(SD) two-forms and triple of connections one form. Later
Ashtekar by considered a canonical transformation on the phase
space of GR,
obtained the SD formulation of GR given by Pleba\'nski.\\
Thus, it was  realized by Capovilla, Dell and
Jacobson\cite{Capovilla}, that the two-form fields of Pleba\'nski
formulation can be integrated out to obtain a pure connection
formulation of GR, where the only dynamical field is the SD
connection  one-form, becoming the main structure used to describe
the gravitational field. The result is that gravity can be
reformulated as a pure connection diffeomorphism invariant gauge theory.\\
Recently Krasnov\cite{KrasnovFF} has shown  that there is not a
single diffeomorphism invariant gauge theory that shares the same
key properties with GR, as they have the same  number of
propagating degrees of freedom (DoF), but an infinite parameter
class of them. Even more, the fundamental scale is set not by the
Newton's constant, which doesn't appear in the original
formulation of the theory at all, but rather by the radius of
curvature of the background that is used to expand the theory
around. Among these diffeomorphism invariant gauge theories,
Gonz\'alez and Montesinos\cite{Montesinos} had introduced pure
gauge connection family of actions that contains not only as the
gauge group the Lorentz group but those that contains a product of
the local Lorentz group and an internal
gauge group, considering vanishing and non-vanishing cosmological constant term.\\
One of the most interesting features given by gauge formulations
is that they are suitable for the incorporation of supersymmetry,
giving rise to the supersymmetric gauge theories. Self-dual and
anti-self-dual formulations of supergravity where considered by
Jacobson\cite{TedJacobson}, and further developed by Capovilla,
Dell and Jacobson\cite{Capovilla}. Later, at the beginning  of the
90's, where considered supergravity self-dual models  in the
Atiyah-Ward space-time\cite{Ketov}. Also, MacDowell-Mansouri
self-dual supergravity models in 3+1 dimensions with $OSp(4|1)$
gauge fields where considered by H.
Garc\'ia-Compe\'an et. al.\cite{Obregon} \\
In this paper, we introduced  a constrained gauge connection
action for gravity in the Lorentzian signature case that  can be
easily generalized to obtain supergravity $N=1$ without the needed
to introduce new dynamical fields. The constrains that will be
used imply algebraic relations, of the Pleba\'nski's type, over
the dynamical fields and provides to the equations of motion the
correct shape needed in each case, for gravity and supergravity
respectively. The organization of the paper is as follows. In
Section 2 We present a brief introduction to BF theory that will
be needed to introduce our proposal for a constrained pure
connection action. In Section 3 We show that SD Einstein spaces
are obtained by considering anti-self-dual (ASD)  connections.
Section 4 We show that supergravity $N=1$ with cosmological
constant sector is obtained by considering ASD $OSp(1|2)$-valued
connection. We conclude with a brief discussion.

%%%%%%%%%%%%%%%%%%%%%%%%%%%%%%%%%%%%%%%%%%%%%%%%%%%%%%%%%%%%%%%%%%%%%%%%%%%%%%%%%%%%%%%%%%%%%%%%%%%%%%%%%%%%%%%%%%%%%%%%%%%%%%%%%%%%%%%%%%%%%%%%%%%%%%%%
%%%%%%%%%%%%%%%%%%%%%%%%%%%%%%%%%%%%%%%%%%%%%%%%%%%%%%%%%%%%%%%%%%%%%%%%%%%%%%%%%%%%%%%%%%%%%%%%%%%%%%%%%%%%%%%%%%%%%%%%%%%%%%%%%%%%%%%%%%%%%%%%%%%%%%%%
%%%%%%%%%%%%%%%%%%%%%%%%%%%%%%%%%%%%%%%%%%%%%%%%%%%%%%%%%%%%%%%%%%%%%%%%%%%%%%%%%%%%%%%%%%%%%%%%%%%%%%%%%%%%%%%%%%%%%%%%%%%%%%%%%%%%%%%%%%%%%%%%%%%%%%%%
%%%%%%%%%%%%%%%%%%%%%%%%%%%%%%%%%%%%%%%%%%%%%%%%%%%%%%%%%%%%%%%%%%%%%%%%%%%%%%%%%%%%%%%%%%%%%%%%%%%%%%%%%%%%%%%%%%%%%%%%%%%%%%%%%%%%%%%%%%%%%%%%%%%%%%%%

\section{ From BF Action to Pure Connection Action}
In this section, we briefly describe some features about
Pleba\'nski theory as a constrained $BF$ theory that will be
needed for the description of our work, for more details the
reader is refereed to
\cite{Plebanski}\cite{Baez-bf}\cite{KrasnovBF}\cite{Freidel-Krasnov-Puzio}
and the references therein. We consider, in general, a principal
fiber bundle $P$ over the four dimensional spacetime manifold
$\mathcal{M}$ with a Lie group $G$ as an internal group, whose Lie
algebra $\textbf{g}$ is equipped with a non degenerate bilinear
invariant form, the Cartan-Killing form $\kappa$, and a $g$-valued
connection $A$ which defines a curvature $F(A)$. It is important
to note that $\mathcal{M}$ is a manifold without a metric
structure  on it and it will be constructed later by means of pure
algebraic relations over
the  basic objects of the theory.\\
In order to construct an action for gravity for the $BF$ theory it
is needed to introduce the two-form Lie algebra valued fields $B$,
 Lagrange multiplier $\Phi$ which is a Lie valued two-form
automorphism whose effect is to constraint the DoF of the $B$
fields and in general without any restriction over it. So that to
obtain the right DoF from the $\Phi$ field we need to put some
constraints on it, to do so, we have two options, one is to impose
the required restrictions by hand, or by means of Lagrange
multipliers, we will use the second option  and then introduce the
four-form fields $\rho_{1}$ and Lie algebra valued fields
$\rho_{2}$ (the  restriction associated to $\rho_{2}$ is
considered in \cite{RamirezRosales} where it was used to construct
a SUSY $N=1$ extension of BF theory). We consider the action
\begin{equation}    \label{eq: BF fundamental}
S_{BF}[A, B, \Phi, \rho_{1}, \rho_{2} ]=\int_{\mathcal{M}} Tr
\bigg( B\wedge F -\frac{1}{2}\  \Phi(B)\wedge B+
\rho_{1}(\Phi-\lambda)+\rho_{2}\Phi \bigg)
\end{equation}
as a constraint $BF$ action for gravity, where $\lambda$ is a
constant, proportional to the cosmological constant (the case
without cosmological constant could be derived if it is considered
$\lambda = 0$). We have to note that we are considering as
internal groups only $SU(2)$ and the supersymmetric extension
$OSp(1|2)$.\\
Let us consider as the generators for Lie algebra $\textbf{g}$ the
set $\{ t_{\alpha}, t_{\beta}, t_{\gamma}, \ldots \}$ where
$[t_{\alpha}, t_{\beta}]=f_{\alpha \beta}^{\ \ \gamma}t_{\gamma}$
, the Cartan-Killing form
$\kappa_{\alpha\beta}=Tr(t_{\alpha}t_{\beta})$, the structure
constants  $f_{\alpha \beta
\gamma}=Tr(t_{\alpha}t_{\beta}t_{\gamma})$ and
$\Phi(B)=\Phi^{\alpha \beta}B_{\alpha}t_{\beta}$. We have to note
that if  we consider superalgebras, the brackets $[ \ ,\ ]$ has to
be changed  by the graded brackets $[ \ ,\ \}$ and the trace,
$Tr$, is changed by supertrace $STr$ in  (\ref{eq: BF
fundamental}).\\
The trace for three generators is valid at last for the internal
group and supergroup considered in this work
\cite{Azcarraga}\cite{Gupta}. Then the action takes the form
\begin{equation}   \label{eq: BF fundamental 2}
S_{BF}[A, B, \Phi, \rho_{1}, \rho_{2} ]=\int_{\mathcal{M}}
B^{\alpha}\wedge F_{\alpha} -\frac{1}{2}\
\Phi^{\alpha\beta}B_{\alpha}\wedge B_{\beta}+
\rho_{1}(\Phi^{\alpha}_{\
\alpha}-\lambda)+\rho_{2}^{\alpha}\Phi^{\beta \gamma}f_{\alpha
\beta \gamma}.
\end{equation}
We have to note that, without loss of generality, we have not
taken into account constant factors that can be singlet out the
action and/or can be absorbed into the Lagrange multipliers. From
(\ref{eq: BF fundamental 2}) we observe that the equations of
motion relative to $\rho_{2}$ and $\rho_{1}$ imply that the $\Phi$
tensor is symmetric and its trace is proportional to the
cosmological constant, respectively. So the action reads
\begin{equation}
S_{BF}[A, B, \Phi ]=\int_{\mathcal{M}} B^{\alpha}\wedge F_{\alpha}
- \frac{1}{2}\ \Phi^{T \alpha\beta}B_{\alpha}\wedge B_{\beta} -
\frac{\lambda}{2} \ B^{\alpha}\wedge B_{\alpha}.
\end{equation}
on which the superindex on $\Phi^T$ is related to the traceless
part of the tensor field. The  last action is how usually it is
presented the $BF$ action for gravity, for internal symmetries as
$SU(2)$ for complex or real fields or in the $SO(3,1)$ with real
fields and Immirzi parameter
\cite{SmolinSpeziale}\cite{EfrainRojas}. Finally, The equations of
motion coming from the last action are
\begin{eqnarray}
 \delta_{A} S_{BF}=0        &\Rightarrow & DB=0 \\
 \delta_{B} S_{BF}=0        &\Rightarrow & F=\Phi^{T}{B}+\lambda B \\
 \delta_{\Phi} S_{BF}=0     &\Rightarrow & (B\wedge B)^{T}=0
\end{eqnarray}
where $D$ is the covariant derivative and it is defined as usual
$D=d\quad+[A,\quad]$. The third equation is known in the
literature as the simplicial or Pleba\'nski constraint
\cite{Plebanski}\cite{Freidel-Krasnov-Puzio}, which implies that
the B-field is a simple two-form of the tetrad field, $B\propto
e\wedge e$, and it was introduced by Pleba\'{n}ski at the middles
of the seventies where he considered as the internal group
$G=SL(2,C)\bigotimes\overline{SL(2,C)}$, later it was considered
only the one quiral part, the complexification of the gauge group
$G=SU(2)$, the self-dual(SD) or anti-self-dual(ASD) description.
Once the explicit form of the B-field is given, the first and the
second equations of motion imply the zero
torsion condition, $A=A(e)$, and the Einstein field equations, respectively.\\
%The  condition associated to $\rho_{2}$ could be thought as an
%unnecessary restriction at the non-supersymmetric case due to the
%fact that in this case  the antisymmetric part of $\Phi$ does not
%play any important role at last at the  Lagrangian level, but  as
%first noted by \cite{Ramirez Rosales}, it plays an important role
%in the supersymmetric extension, as we shall see.\\
The action is invariant under local gauge transformation
\begin{equation}
\delta_{\alpha}^{G} B =-[\alpha, B] \qquad \delta_{\alpha}^{G} A
=-D \alpha \qquad \delta_{\alpha}^{G} \Phi =-[\alpha,
\Phi]^{\textbf{g}\otimes\textbf{g}}\qquad \delta_{\alpha}^{G}
\rho_{2} =-[\alpha, \rho_{2}]
\end{equation}
where $\alpha$ are Lie algebra-valued scalars (gauge parameters)
and the symbol $[\quad,\quad]^{\textbf{g}\otimes\textbf{g}}$ is
due that  $\Phi$ is a Lie algebra valued bivector so the gauge
transformation must be taken on each vector index. But even more,
the action is invariant under the Kalb-Ramond(shift) symmetry
\cite{Horowitz}\cite{Freidel-Speziale}
\begin{equation}         \label{eq: Kalb-Ramond symmetry}
\delta_{C} B =-DC \qquad \delta_{C} A =\Phi\cdot C
\end{equation}
where C are Lie algebra one-form transformation parameters. If we
defined $C=i_{v}B$, then the diffeomorphisms plus field dependent
gauge transformations are found on-shell, but not only the
transformation of the basic fields $B$ and $A$, but we can find
the rules of transformation over the rest fields $\Phi$,
$\rho_{1}$ and $\rho_{2}$ implying that the total symmetry group
is the semi-direct product of these two groups.\\
Finally, let us  construct  a  pure connection action that we will
consider in this work. Inspired by the equation of motion for the
B-field \cite{Capovilla}\cite{Smolin}, $F=\Phi\cdot B$, consider
det\ $ \Phi \neq 0$ so $B=\Phi^{-1}F$, then the action (\ref{eq:
BF fundamental}) is rewritten as
\begin{equation}
S_{BF}[A, \Phi, \Psi, \rho_{1}, \rho_{2} ]=\int_{\mathcal{M}} Tr
\bigg( \frac{1}{2}\ \Psi(F)\wedge F+
\rho_{1}(\Phi-\lambda)+\rho_{2}\Phi \bigg)
\end{equation}
where we have defined $\Psi=\Phi^{-1}$. But the relations on
$\Phi$ are algebraic so we have to rewrite the restrictions over
$\Psi$ in an algebraic manner. Then our proposal is to take the
action
\begin{equation}    \label{eq: CDJ fundamental}
S_{FF}[A, \Psi, \rho_{1}, \rho_{2} ]=\int_{\mathcal{M}} Tr \bigg(
 \Psi(F)\wedge F+ \rho_{1}\Psi+\rho_{2}\Psi \bigg)
\end{equation}
as topological pure gauge action for gravity independent of the
$BF$ functional. We consider for simplicity  $\Psi$ as a traceless
object because opposite to the $BF$ theory, the cosmological
constant does not enter into the theory by this constraint and
even more, we can observe that if we consider the trace term a
constant different from zero, it only contributes as a boundary
term in the form of the Chern-Simmons functional with no effect at
the classical level which is the level presented in this work.\\
We have to note that it inherites the same gauge and
diffeomorphism symmetry as in the $BF$ theory that could be seen
through the shift symmetry shown in equation (\ref{eq: Kalb-Ramond symmetry}), and %%%%%%%%%%%%%%%%\begin{equation}\delta_{C}= i_{v}F\end{equation}
has the same topological spirit, in the sense, that the metric
doesn't appear explicitly and it can be seen as a perturbation of
a topological field theory, for  if we consider
$\Psi^{\alpha\beta}\approx \delta^{\alpha\beta}$, the
Lagrangian becomes a total derivative, the second Chern Class.\\
In the next two sections we will show that the equation of motion
coming from this action, considering complex field as well as the
gauge group $SU(2)$ and $OSp(2|1)$, are ASD solutions of the
Einstein equations, for gravity and supergravity respectively,
where the zero torsion and supertorsion condition,    comes by
means of the Bianchi identity. Through the work we have labelled
$su(2)$ Lie algebra indices by the middle of the Latin alphabet
lowercase letters $\{ i, j, k, \ldots \}$, $so(1,3)$ Lie algebra
indices by the beginning of the Latin alphabet  lowercase letters
$\{a, b, c, \ldots \}$, capital Latin letters for ASD spinorial
indices $\{A, B, C, \ldots \}$, overdotted capital Latin letters
for SD spinorial indices $\{\dot{A}, \dot{B}, \dot{C}, \ldots \}$,
Lie superalgebra $OSp(1|2)$ indices by the end of the Latin
alphabet lowercase letters $\{p, q, r \ldots \}$ and Greek
alphabet letters for space-time indices $\{\mu, \nu, \rho, \ldots
\}$. Finally we define
$G^{(\alpha\beta)}=G^{\alpha\beta}+G^{\beta\alpha}$ and
$G^{[\alpha\beta]}=G^{\alpha\beta}-G^{\beta\alpha}$.

%%%%%%%%%%%%%%%%%%%%%%%%%%%%%%%%%%%%%%%%%%%%%%%%%%%%%%%%%%%%%%%%%%%%%%%%%%%%%%%%%%%%%%%%%%%%%%%%%%%%%%%%%%%%%%%%%%%%%%%%%%%%%%%%%%%%%%%%%%%%%%%%%%%%%%%%
%%%%%%%%%%%%%%%%%%%%%%%%%%%%%%%%%%%%%%%%%%%%%%%%%%%%%%%%%%%%%%%%%%%%%%%%%%%%%%%%%%%%%%%%%%%%%%%%%%%%%%%%%%%%%%%%%%%%%%%%%%%%%%%%%%%%%%%%%%%%%%%%%%%%%%%%
%%%%%%%%%%%%%%%%%%%%%%%%%%%%%%%%%%%%%%%%%%%%%%%%%%%%%%%%%%%%%%%%%%%%%%%%%%%%%%%%%%%%%%%%%%%%%%%%%%%%%%%%%%%%%%%%%%%%%%%%%%%%%%%%%%%%%%%%%%%%%%%%%%%%%%%%
%%%%%%%%%%%%%%%%%%%%%%%%%%%%%%%%%%%%%%%%%%%%%%%%%%%%%%%%%%%%%%%%%%%%%%%%%%%%%%%%%%%%%%%%%%%%%%%%%%%%%%%%%%%%%%%%%%%%%%%%%%%%%%%%%%%%%%%%%%%%%%%%%%%%%%%%
\section{ASD formulation for gravity}
Let us take the Lie group $SU(2)$ as our gauge group, this group
is semisimple so  Cartan-Killing form is non-degenerated. We take
as our spacetime a 4-dimensional globally hyperbolic, oriented
smooth manifold $\mathcal{M}$. Now choose a principal complex
$SU(2)$-bundle $P$ over $M$ which is related to the ASD complex
bundle. Let us take as our fundamental dynamical field a $SU(2)$
complex connection $A=A_{i}t^{i}$ ($i=1,2,3$) where $A^{i}$ is the
ASD part of the real $SO(3,1)$ connection $A^{ab}$ ($a, b=0, 1, 2,
3$)
\begin{equation}       \label{eq: Antiselfdual projector}
A^{i}=\Pi^{(-)0i}_{\ \ \ \ \  ab} A^{ab} \qquad \textrm{where}
\quad \Pi^{(-)0i}_{\ \ \ \ \ ab}=\frac{1}{4}\ \bigg(\eta^{0i}_{\ \
ab}+i\epsilon^{0i}_{\ \  ab} \bigg),
\end{equation}
from which we have defined
$\eta_{ab,cd}=\eta_{ac}\eta_{bd}-\eta_{ad}\eta_{bc}$ and
$\eta=diag(-1, 1, 1, 1)$, the Minkowski metric. The Levi-Civita
tensor is taken as $\epsilon^{0123}=1$,
$\epsilon^{0ijk}=\epsilon^{ijk}$ and $\epsilon^{123}=1$. The
generators of the $su(2)$ Lie algebra in the adjoint
representation satisfy
\begin{equation}
[t_{i},t_{j}]=f_{ij}^{\ \ k}t_{k}=2i\epsilon_{ij}^{\ \  k}t_{k}.
\end{equation}
The Cartan-Killing form is calculated directly by the trace over
the Lie algebra generators and we observe that is proportional to
the Kronecker delta, then as the internal metric is defined upon
an overall constant we consider $\kappa_{ij}=\delta_{ij}$. %\footnote{Formally the Cartan-Killing form is defined upon a constant factor, i.e.$\kappa_{ij}=Tr(t_{i}t_{j})/I_{ad}$ to which $I_{ad}$ is a constant factor that can be single out of the action and absorbedby the Lagrange multipliers.}
The field strength is then an ASD two form given by
\begin{equation} \label{eq: Bosonic field strength}
F_{i}=dA_{i}+i\epsilon_{i}^{\ jk}A^{j}\wedge A^{k}.
\end{equation}
Our next step is to define the zero-form field $\Psi$ as a
bivector Lie algebra valued field, $\Psi=\Psi^{ij}t_{i}t_{j}$, and
else, define the action over the two-from field strength as
$\Psi(F)=\Psi^{ij}F_{i}t_{j}$. At this point we have all the
necessary ingredients needed in (\ref{eq: CDJ fundamental}) so in
this case the action is written as
\begin{equation}
S_{FF}[A, \Psi, \rho_{1}, \rho_{2} ]=\int_{\mathcal{M}}
 \Psi^{ij} F_{i}\wedge F_{j}+ \rho_{1}\Psi^{i}_{\ i}+\rho_{2}^{i}\Psi^{jk}\epsilon_{ijk}.
\end{equation}
Let us now calculate the equations of motion, we consider first
those that put constraints  in the shape of $\Psi$ field. From
$\rho_{2}$ we obtain that $\Psi$ is completely symmetric object
and from $\rho_{1}$ we obtain that $\Psi$ is traceless. These
equations of motion are algebraic so the action is equivalent to
\begin{equation}
S_{FF}[A, \Psi]=\int_{\mathcal{M}}
 \Psi_{ij}^{T} F^{i}\wedge F^{j}.
\end{equation}
From the variation of $\Psi^{T}$ field, we obtain
\begin{equation}           \label{eq: simplicialFF}
(F^{i}\wedge F^{j})^{T}=F^{i}\wedge F^{j}-\frac{1}{3}\
\delta^{ij}F^{k}\wedge F_{k}=0.
\end{equation}
The last equation is known as the instanton equation
\cite{Torre}\cite{JanisPorter}, it is a simplicial constraint
similar to the Pleba\'nski constraint
 and it implies is that $F$ is an ASD simple two form
\begin{equation}     \label{eq: f equaltetrad field}
F^{i}=\lambda \Pi^{(-)0i}_{\ \ \ \ ab}\Sigma^{ab}=\lambda
\Sigma^{i} \qquad \textrm{where} \quad \Sigma^{ab}=e^{a}\wedge
e^{b}
\end{equation}
in which $\lambda$ is a constant proportional to the cosmological
constant, by dimensional consistency,  and it appears due to the
fact that the algebraic equation is defined upon a constant term.
Besides if we apply the covariant derivative in both sides of the
equation (\ref{eq: f equaltetrad field}) and by means of the
Bianchi identitiy,  we obtain
\begin{equation}
DF^{i}=D\big( \lambda \Pi^{(-)0i}_{\ \ \ \
ab}\Sigma^{ab}\big)=\lambda D\Sigma^{i}\Rightarrow D\Sigma^{i} =0
\end{equation}
which is the zero torsion  condition and it implies that the ASD
connection can be written as a function of the tetrad field,
$A=A(e)$. So  if we demand that the tetrad field is nondegenerate
then (\ref{eq: f equaltetrad field}) leads to conformally ASD
Einstein spaces where the field strength has SD Weyl tensor
\cite{SmolinSpeziale}\cite{Smolin}\cite{Torre}.\\
The equation of motion for the connection $A$, gives
\begin{equation}             \label{eq: PsiF covconstant}
D(\Psi_{ij}^{T}F^{j})=0
\end{equation}
which implies that $\Psi_{ij}^{T}\ F^{j}$ is covariantly constant.
The equation (\ref{eq: PsiF covconstant}) is a differential
equation that gives a dynamical behavior to the $\Psi$ field, so
it is interesting to note that it gives us an additional algebraic
equation by  applying a covariant derivative over itself, called
integrability condition. For simplicity instead of taking only the
traceless part of $\Psi$, we consider the complete tensor field
$\Psi$, and then put the traceless and antisymmetric restriction
over it
\begin{equation}   \label{eq: Derivative PsiF covconstant}
D(D(\Psi_{ij}F^{j}))=0 \Rightarrow \epsilon^{i}_{\ jk}\ \Psi_{il}\
F^{l}\wedge F^{j}=\epsilon^{i}_{\ jk}\ \Psi^{T}_{il}\ F^{l}\wedge
F^{j}=0
\end{equation}
and by the equation of motion imposed by $\rho_{2}$, as well as
the not degeneracy of the tetrad field, it implies that the
equation (\ref{eq: Derivative PsiF covconstant}) is an
identity,i.e. it does not put any new constraint into the theory.
In general, the solution for the integrability condition is not
trivial, as we shall see in the supersymmetric case, and put
constraints over $\Psi$ that has to be taken into account in the
theory, fortunately for us,  the solution is very trivial in the
bosonic case. Finally, we observe that we can obtain ASD Einstein
gravity from the action (\ref{eq: CDJ
fundamental}) by the use of algebraic equations and the Bianchi identity.\\
The $FF$ action inherites the symmetry coming from the BF action,
i.e., the action is invariant under the local gauge transformation
\begin{equation}
\delta_{\alpha} A^{i} =-D \alpha^{i} \qquad \delta_{\alpha}
\Psi^{ij} =-2i\epsilon_{kl}^{\ \ i} \alpha^{k}
\Psi^{lj}-2i\epsilon_{kl}^{\ \ j} \alpha^{k} \Psi^{il} \qquad
\delta_{\alpha} \rho_{2}^{i} =-2i\epsilon_{kl}^{\ \
i}\alpha^{k}\rho_{2}^{l}
\end{equation}
where $\alpha$ are $su(2)$-valued scalars, and the shift symmetry
\begin{equation}
 \delta_{C} A =-2C
\end{equation}
for if C are $su(2)$ one-form transformation parameters defined as
$C=i_{v}F=v^{\nu}F_{\mu\nu}dx^{\mu}$, then  diffeomorphisms plus
field dependent gauge transformations of all the fields are given
\begin{eqnarray}
\nonumber  \delta_{v}A_{\nu}^{\ i}      &=& v^{\mu}\partial_{\mu} A_{\nu}^{\ i}+\partial_{\nu}v^{\mu} A_{\mu}^{\ i}-D_{\nu}{\alpha}^{i} \\
\nonumber  \delta_{v}\Psi^{ij}          &=& v^{\mu}\partial_{\mu}\Psi^{ij}-\delta_{\alpha}\Psi^{ij} \\
\nonumber  \delta_{v} \rho_{1}          &=& D_{\mu}(v^{\mu} \rho_{1}) \\
\nonumber  \delta_{v} \rho_{2}^{\ i}    &=& \partial_{\mu}(v^{\mu}
\rho_{2}^{\ i})-\delta_{\alpha}\rho_{2}^{\ i}
\end{eqnarray}
and $\delta_{\alpha}$ are $SU(2)$ transformations with parameters
$\alpha^{i}=v^{\mu}A_{\mu}^{\ i}$ and the four-forms $\rho_{1}$,
$\rho_{2}^{\ i}$ transform as invariant densities. At this point
we have to note that the rules of transformation can be found
without the use of the equation of motion, i.e., they are obtained
off-shell oppositive to the tetrad field case where
transformations can be found by means of the equation of motion
(\ref{eq: f equaltetrad field}).\\
Finally we end this section by pointing some remarks that will be
useful in the next section, we have defined ASD  fields as $su(2)$
Lie valued complex fields $v^{i}$ which can be related to
$so(3,1)$ Lie valued real fields $v^{ab}$ by the ASD projector
$\Pi^{(-)}$ (\ref{eq: Antiselfdual projector}), but originally,
the action of Pleba\'nski was formulated in spinorial notation and
we can recover the original equations by considering
$v^{i}=\frac{1}{2}\ (\sigma^{i})_{A}^{\ B}v_{B}^{\ A}$, where
$(\sigma^{i})$ are the Pauli matrices. On the other hand, for the
Lorentz group the corresponding embedding is into $SL(2,C)\otimes
\overline{SL(2,C)}$, given for the fundamental representation by
$v_{A\dot{B}}=v_{a}(\sigma^{a})_{A\dot{B}}$, where
$(\sigma^{0})_{A\dot{B}}$ is the identity matrix and
$(\sigma^{i})_{A\dot{B}}$ are the Pauli matrices. With these
conventions, the adjoint representation of $SO(3,1)$ decomposes
as,
\begin{equation}
v_{ab}=(\sigma^{ab})_{A}^{\ B}v_{B}^{\ A}
+(\overline{\sigma}^{ab})_{\ \dot{B}}^{ \dot{A}}v_{\ \dot{A}}^{
\dot{B}}
\end{equation}
where $\sigma^{ab}=\frac{1}{4}\
[\sigma^{a},\overline{\sigma}^{b}]$ and
$\overline{\sigma}^{ab}=\frac{1}{4}\
[\overline{\sigma}^{a},\sigma^{b}]$, satisfy $\epsilon^{ab}_{\ \
cd}\sigma^{cd}=-2i\sigma^{ab}$ and $\epsilon^{ab}_{\ \
cd}\overline{\sigma}^{cd}=2i\overline{\sigma}^{ab}$. Hence
$v_{AB}$ and $v_{\dot{A}\dot{B}}$ are ASD and SD, respectively.
Then we can give the relations between objects valued in $su(2)$,
$so(3,1)$ and $sl(2,C)$ as follows
\begin{equation} \label{eq: Fundamental relations among Lee algebras}
v_{A}^{\ B}=(\sigma^{i})_{A}^{\ B}v_{i}=\Pi^{(-)0i}_{\ \ \ \ \
ab}(\sigma^{i})_{A}^{\ B}v^{ab}=-\frac{1}{2}\ (\sigma_{ab})_{A}^{\
B}v^{ab}.
\end{equation}
From the last result we can obtain  $u^{i}v_{i}=-\frac{1}{4} \
u^{(-)ab}v_{ab}^{(-)}=-\frac{1}{2}\ u^{AB}v_{AB}$. We can lower
and rise indices by the two dimensional Levi-Civita tensor
$\epsilon$,
\begin{equation}
\epsilon^{AB}\xi_{B}=\xi^{A} \qquad \epsilon_{AB}\xi^{B}=\xi_{A}
\qquad \epsilon^{01}=1.
\end{equation}
Now we can translate the equations of motion (\ref{eq: f
equaltetrad field}) and its covariant derivative, the zero torsion
condition, into the spinor language
\begin{equation}
F^{AB}=\lambda_{1}\ \Sigma^{AB} \qquad ,\qquad D\Sigma^{AB}=0
\end{equation}
where $\Sigma^{AB}=e^{A\dot{A}}\wedge e_{\dot{A}}^{\ B}$
\cite{Capovilla}.

%%%%%%%%%%%%%%%%%%%%%%%%%%%%%%%%%%%%%%%%%%%%%%%%%%%%%%%%%%%%%%%%%%%%%%%%%%%%%%%%%%%%%%%%%%%%%%%%%%%%%%%%%%%%%%%%%%%%%%%%%%%%%%%%%%%%%%%%%%%%%%%%%%%%%%%%
%%%%%%%%%%%%%%%%%%%%%%%%%%%%%%%%%%%%%%%%%%%%%%%%%%%%%%%%%%%%%%%%%%%%%%%%%%%%%%%%%%%%%%%%%%%%%%%%%%%%%%%%%%%%%%%%%%%%%%%%%%%%%%%%%%%%%%%%%%%%%%%%%%%%%%%%
%%%%%%%%%%%%%%%%%%%%%%%%%%%%%%%%%%%%%%%%%%%%%%%%%%%%%%%%%%%%%%%%%%%%%%%%%%%%%%%%%%%%%%%%%%%%%%%%%%%%%%%%%%%%%%%%%%%%%%%%%%%%%%%%%%%%%%%%%%%%%%%%%%%%%%%%
%%%%%%%%%%%%%%%%%%%%%%%%%%%%%%%%%%%%%%%%%%%%%%%%%%%%%%%%%%%%%%%%%%%%%%%%%%%%%%%%%%%%%%%%%%%%%%%%%%%%%%%%%%%%%%%%%%%%%%%%%%%%%%%%%%%%%%%%%%%%%%%%%%%%%%%%

\section{Supersymmetric extension}
The generalization of action (\ref{eq: CDJ fundamental}) for
supergravity $N=1$ is done by making its fundamental fields
transforming under the adjoint representation of $OSp(2|1)$. The
fermionic  (odd) part  is labelled by the spinorial indices and
the bosonic (even) part by $su(2)$ indices, then the connection
one-form is written as $A=A^{p}t_{p}=A^{i}t_{i}+A^{B}t_{B}$,
$\Psi=\Psi^{pq}t_{p}t_{q}$ and $\Psi(F)=\Psi^{pq}F_{q}t_{p}$.\\
The generators of the Lie superalgebra are given by $[t_{p},
t_{q}\}=f_{pq}^{\ \ r}t_{r}$ where $f_{pq}^{\ \ r}$ are the
structure constants and are calculated as follows
\begin{equation}
[t_{i}, t_{j}]=2i\epsilon_{ij}^{\ \ k} \qquad [t_{i},
t_{A}]=-(\sigma_{i})_{A}^{\ B}t_{B} \qquad \{ t_{A}, t_{B}
\}=(\sigma^{i})_{AB}t_{i}.
\end{equation}
The Cartan-Killing form is calculated straightforward
\begin{equation}
 \kappa_{pq}=\left(%
\begin{array}{cc}
  \delta_{ij} & 0 \\
  0           & \epsilon_{AB} \\
\end{array}%
\right).
\end{equation}
The  bosonic and fermionic sectors of the field strength  are
\begin{equation}
  F^{i} = R^{i}-\frac{1}{2}\ A (\sigma^{i})A \qquad ,\qquad
  F^{B} = d A^{B}-(\sigma_{i})_{C}^{\ B}A^{i}\wedge A^{C}
\end{equation}
where $R^{i}$ is the bosonic ASD field strength (\ref{eq: Bosonic
field strength}). Let us defined the covariant derivative for Lie
superalgebra valued fields as $\nabla \xi=\nabla
\xi^{i}t_{i}+\nabla \xi^{B}t_{B}$, where
\begin{displaymath}
  \nabla \xi^{i} = D\xi^{i}-(\sigma^{i})_{C}^{\ B} A^{C}\wedge \xi_{B}
  \qquad ,\qquad
  \nabla \xi^{B} = D\xi^{B}+(\sigma_{j})_{C}^{\ B} A^{C}\wedge
\xi^{j}
\end{displaymath}
and $D$ is referred to the usual non supersymmetric covariant
derivative which it is defined in such a way that it ``knows" how
to act over different non supersymmetric Lie algebra valued fields
\begin{displaymath}
   D \zeta^{i}  = d\zeta^{i} +2i\epsilon_{jk}^{\ \ i }A^{j}\wedge
   \zeta^{k}\qquad \quad
   D \zeta^{ab} = d\zeta^{ab}+A^{ac}\wedge \zeta_{c}^{\ b}+A^{bc}\wedge \zeta_{\ c}^{a}
\end{displaymath}
\begin{displaymath}
   D \zeta^{B}  = d\zeta^{B} +A^{BC}\wedge \zeta_{C} \qquad \qquad
   D \zeta^{\dot{B}}  = d\zeta^{\dot{B}} +A^{\dot{B}\dot{C}}\wedge
   \zeta_{\dot{C}}.
\end{displaymath}
Now we have  the basic ingredients needed in the action (\ref{eq:
CDJ fundamental}) which in the supersymmetric case is written as
follows
\begin{equation}    \label{eq: SUSY fundamental}
S_{FF}[A, \Psi, \rho_{1}, \rho_{2} ]=\int_{\mathcal{M}}
 \Psi^{pq} F_{p}\wedge F_{q}+ \rho_{1}\Psi^{p}_{\ p}+\rho_{2}^{p}\Psi^{qr}f_{pqr}.
\end{equation}
Now, from the variation of $\rho_{1}$, we have  that $\Psi$ is
supertraceless, $\Psi^{p}_{\ p}=\Psi^{i}_{\ i}+\Psi^{A}_{\ A}=0$.
From the variation of $\rho_{2}$ we obtain $\Psi^{qr}f_{pqr}=0$
which implies two independent relations
\begin{eqnarray}
  2i\Psi^{ij}\epsilon_{ijk}- \Psi^{AB}(\sigma_{k})_{AB}&=& 0 \\
\label{eq: constraint mixed}
(\Psi^{iA}+\Psi^{Ai})(\sigma_{i})_{AB}=-\Psi_{B\  D}^{\ \
D}+\Psi^{D}_{\ DB}  &=& 0.
\end{eqnarray}
We note that in the supersymmetric formulation, the antisymmetric
part of the bosonic sector does not vanish, instead it implies
that
\begin{displaymath}
\Psi^{[ij]}=-\frac{1}{4}\
[\sigma^{i},\sigma^{j}]_{AB}\Psi^{AB}\quad , \quad
\Psi^{(AB)}=[\sigma^{i},\sigma^{j}]^{AB}\Psi_{ij},
\end{displaymath}
fortunately, even when they both give contributions to the
antisymmetric bosonic sector of $\Psi^{ij}$ and to the symmetric
part of the fermionic sector $\Psi^{AB}$ they are not necessary
because of the symmetry of the wedge product of the field strength
with itself,i.e. $F\wedge F$.\\
Now  the action reads
\begin{equation}
S_{FF}[A, \Psi ]=\int_{\mathcal{M}}
 \Psi^{ij} F_{i}\wedge F_{j}+\Psi^{AB} F_{A}\wedge F_{B}+(\Psi^{iA}+\Psi^{Ai}) F_{i}\wedge F_{A}
\end{equation}
but in order to go further, we have to decompose $\Psi$ into its
irreducible components \cite{Penrose}
\begin{eqnarray}
  \Psi^{ij}              &=& \Psi^{T\ ij}+\frac{1}{3}\ \delta^{ij}\Psi^{k}_{\ k} \\
  \Psi^{AB}              &=& -\frac{1}{2}\ \epsilon^{AB} \Psi^{C}_{\ C}+\frac{1}{2}\ \Psi^{(AB)}\\
  \Psi^{iA}+\Psi^{Ai}    &=&
  -(\sigma^{i})_{BC}\Psi^{(ABC)}+\frac{1}{3}\ (\sigma_{i})_{C}^{\
  A}\Big[-\Psi_{\  \ \ D}^{CD}+\Psi^{D\  C}_{\ D} \Big]
\end{eqnarray}
where $\Psi^{(ABC)}$ and $\Psi^{(AB)}$ are  completely symmetric
tensor fields. As  (\ref{eq: constraint mixed}) is an algebraic
equation and the product $F_{A}\wedge F_{B}$ is antisymmetric, the
action is equivalent to
\begin{equation}
S_{FF}[A, \Psi ]=\int_{\mathcal{M}} \Psi^{T\ ij} F_{i}\wedge
F_{j}+\Psi^{k}_{\ k} \bigg( \frac{1}{3}\ F^{i}\wedge F_{i}
-\frac{1}{2}\  F^{B}\wedge F_{B}\bigg)-\Psi^{(ABC)} F_{AB}\wedge
F_{C}.
\end{equation}
From the last action, let us calculate the equations of motion.
For the bosonic traceless part of $\Psi$, we have  the Pleba\'nski
simplicial constraint
\begin{equation}
(F_{i}\wedge F_{j})^{T}=0 \Rightarrow F_{i}=R_{i}-\frac{1}{2}\
A(\sigma_{i})A=\lambda_{1}\Pi^{(-)0i}_{\ \ \ \ \
ab}\Sigma^{ab}=\lambda_{1}\Sigma_{i}.
\end{equation}
From $\Psi^{(ABC)}$ we obtain the well known CDJ constraint
\cite{Capovilla}, where  the solution is defined, upon a constant
factor, as
\begin{equation}  \label{eq: Solution CDJ constraint}
F_{(AB}\wedge F_{C)}=0 \Rightarrow DA_{C}=F_{C}=\lambda_{2}
(\sigma_{a})_{C\dot{C}} e^{a}\wedge
\overline{\varphi}^{\dot{C}}=\lambda_{2}\Pi_{C}^{\
ab}\Sigma_{ab}=\lambda_{2}\Sigma_{C}
\end{equation}
where $\varphi$ is a spin $3/2$ one-form field which at the
quantum level, together with reality conditions
\cite{Capovilla}\cite{RamirezRosales}, is known as the gravitino
field, $\overline{\varphi}$ is its complex conjugate. Thus we have
defined the spinorial proyector $\Pi_{C}^{\
ab}=(\sigma^{a})_{C\dot{C}}\overline{\varphi}^{b\dot{C}}$. \\
For the equation of motion for the connection $A$, we proceed as
in the last section and, at first, we calculate the most general
form of the equation of motion coming from the action (\ref{eq:
SUSY fundamental}) and then put the constraints over it. Then we
have for the connection
\begin{equation}  \label{eq: Superderivative PsiF covconstant}
\nabla[(\Psi^{pq}+(-1)^{|p||q|}\Psi^{qp})F_{q}]=0
\end{equation}
where the symbol $|\cdot|$ is zero if the object is bosonic and
one if it is  a fermionic index. First let us apply a
supercovariant derivative to (\ref{eq: Superderivative PsiF
covconstant}) in order to obtain the integrability condition
\begin{equation}
\nabla\big[\nabla[(\Psi^{pq}+(-1)^{|p||q|}\Psi^{qp})F_{q}]
\big]=(\Psi^{sq}+(-1)^{|s||q|}\Psi^{qs})F^{r}\wedge F_{q}f_{rs}^{\
\ p} =0,
\end{equation}
then by the use of the simplicial constraints, symmetry properties
of $\Psi$ (and a little bit of algebra),  we obtain
\begin{equation}
\rho_{2}^{D}\Psi^{A}_{\ A}=0,
\end{equation}
but we observe, by consistency with the equation of motion
(\ref{eq: constraint mixed}), that $\rho_{2}^{D}\neq 0$ (otherwise
(\ref{eq: constraint mixed}) couldn't exists), and so $\Psi^{A}_{\
A}=0$, implying by supertraceless condition, that $\Psi^{i}_{\ i}$
is zero too. As we can observe, in the supersymmetric case we
obtain, from the integrability condition, a new constraint over
the $\Psi$
field, which implies that  each trace term in $\Psi$ vanish independently.\\
Finally, by the use of the Bianchi's identity  $\nabla F=\nabla
F^{i}t_{i}+\nabla F^{A}t_{A}=0$ we obtain
\begin{equation}
  D \Sigma_{i} = \frac{\lambda_{2}}{\lambda_{1}}\ A^{B}\wedge \Sigma_{C}(\sigma_{i})_{B}^{\ C}
  \qquad , \qquad
  D \Sigma_{A} =
  -\frac{\lambda_{1}}{\lambda_{2}}\ A_{B}\wedge
  \Sigma_{A}^{\ B}.
\end{equation}
We observe that the solution of the CDJ simplicial constraint
(\ref{eq: Solution CDJ constraint}) force us to introduce into the
game, the complex conjugate quiral part of the ASD sector, i.e.,
the objects with dotted indices, then for a correct description of
the theory we have to translate the equations from the $SU(2)$
complex valued fields to the $SL(2,C)\otimes \overline{SL(2,C)}$
language, given by the relations  at the end of the last section
(\ref{eq: Fundamental relations among Lee algebras}). But even
more it is also necessary if we wish to compare our result with
those that we can find in the literature. Then let us identify
$A_{C}=\sqrt \Lambda \ \varphi_{C}$, $\lambda_{1}=-4\Lambda^{2}$
and $\lambda_{2}=-i\Lambda^{3/2}$ and by (\ref{eq: Fundamental
relations among Lee algebras}), the equations of motion read
\begin{eqnarray}
\nonumber  (\sigma_{ab})_{A}^{\ B} \bigg[R^{ab}+4\Lambda^{2}\Sigma^{ab}\bigg]        &=& -2\Lambda  \varphi_{A}\wedge\varphi^{B}  \\
\nonumber  De^{a}                                                                          &=& -\frac{i}{2}\ \overline{\varphi}(\sigma^{a})\varphi\\
\nonumber  e^{a}(\overline{\sigma}_{a})^{\dot{A}A}\wedge D\varphi_{A}                      &=& -2i\Lambda (\overline{\sigma}_{ab})^{\dot{A}}_{\ \dot{B}} \Sigma^{ab}\wedge \overline{\varphi}^{\dot{B}} \\
\nonumber  e^{a}(\overline{\sigma}_{a})^{\dot{A}A}\wedge
D\overline{\varphi}_{\dot{A}}
  &=& +2i\Lambda (\sigma_{ab})_{B}^{\ A} \Sigma^{ab}\wedge
  \varphi^B.
\end{eqnarray}
These equations can be  compared to the usual equations of motion
for supergravity N=1  in ASD formulation with cosmological
constant sector
\cite{TedJacobson}\cite{Wess-Bagger}\cite{Townsend}, we observe
that the first is a solution for supergravity which is the
supersymmetric extension of (\ref{eq: f equaltetrad field}), the
second is the zero supertorsion condition and the last two gives
the dynamical behavior of the gravitino field related by complex
conjugation.\\
The rules of transformation for the fields are straightforward
calculated based on the  previous section results, the shift
symmetry $\delta_{C}A^{p}=v^{\mu}F_{\nu\mu}^{\ \ p}dx^{\nu}$
implies off-shell transformations
\begin{eqnarray}
\nonumber  \delta_{v}A_{\nu}^{\ p}      &=& v^{\mu}\partial_{\mu} A_{\nu}^{\ p}+\partial_{\nu}v^{\mu} A_{\mu}^{\ p}-\nabla_{\nu}{\alpha}^{p} \\
\nonumber  \delta_{v}\Psi^{pq}          &=& v^{\mu}\partial_{\mu}\Psi^{pq}-\delta_{\alpha}\Psi^{pq} \\
\nonumber  \delta_{v} \rho_{1}          &=& D_{\mu}(v^{\mu} \rho_{1}) \\
\nonumber  \delta_{v} \rho_{2}^{\ p}    &=& \partial_{\mu}(v^{\mu}
\rho_{2}^{\ p})-\delta_{\alpha}\rho_{2}^{\ p}
\end{eqnarray}
where $\delta_{\alpha}$ are $Osp(1|2)$ transformations with
parameters $\alpha^{p}=v^{\mu}A_{\mu}^{\ p}$ and the four-forms
$\rho_{1}$, $\rho_{2}^{\ p}$ transform as invariant densities and
the rules for the tetrad field transformations  can
be found off-shell.\\

%%%%%%%%%%%%%%%%%%%%%%%%%%%%%%%%%%%%%%%%%%%%%%%%%%%%%%%%%%%%%%%%%%%%%%%%%%%%%%%%%%%%%%%%%%%%%%%%%%%%%%%%%%%%%%%%%%%%%%%%%%%%%%%%%%%%%%%%%%%%%%%%%%%%%%%%
%%%%%%%%%%%%%%%%%%%%%%%%%%%%%%%%%%%%%%%%%%%%%%%%%%%%%%%%%%%%%%%%%%%%%%%%%%%%%%%%%%%%%%%%%%%%%%%%%%%%%%%%%%%%%%%%%%%%%%%%%%%%%%%%%%%%%%%%%%%%%%%%%%%%%%%%
%%%%%%%%%%%%%%%%%%%%%%%%%%%%%%%%%%%%%%%%%%%%%%%%%%%%%%%%%%%%%%%%%%%%%%%%%%%%%%%%%%%%%%%%%%%%%%%%%%%%%%%%%%%%%%%%%%%%%%%%%%%%%%%%%%%%%%%%%%%%%%%%%%%%%%%%
%%%%%%%%%%%%%%%%%%%%%%%%%%%%%%%%%%%%%%%%%%%%%%%%%%%%%%%%%%%%%%%%%%%%%%%%%%%%%%%%%%%%%%%%%%%%%%%%%%%%%%%%%%%%%%%%%%%%%%%%%%%%%%%%%%%%%%%%%%%%%%%%%%%%%%%%
\section{Conclusions and Outlooks}

As was shown by Krasnov and Montesinos,  there are a family of
diffeomorphism invariant pure connection gauge theories that share
the same key properties with General Relativity, then among those
elements of the family, we introduce a simple one (\ref{eq: CDJ
fundamental}). As we shown by the Kalb-Ramond symmetry, is gauge
and diffeomorphism invariant, as expected. But even more, we
showed how to obtain SD Einstein spaces by consider ASD complex
$su(2)$ valued fields, and supergravity $N=1$ with cosmological
sector by considering ASD complex $OSp(1|2)$ fields. We observed,
in both cases, that half of the equation of motions needed in the
theory came by the variation of the action and the left half came
by the use of the Bianchi identity. We also had that we haven't
put constraints over the $\Psi$ field, as it is usual done,
instead this field has no intrinsic properties and it is
constrained, in its shape, by the constrains given in the action,
and by integrability conditions. As we showed the trace of the
$\Psi$ field is not longer related to the cosmological constant
and in both cases, pure bosonic and supersymmetric cases,
vanishes. Finally, the appearance of the cosmological constant  is
due to algebraic relations and dimensional consistency, and it has
no relation with the trace part, as in the $BF$ case.\\
It would be interesting to consider the canonical analysis of the
class to verify that the theory has two degrees of freedom per
spacetime point and to compare to those that could be found in the
literature, as special case, how different is from the Ashtekar
formulation. All of these, in order to obtain the quantization of
the theory proposed. \\
We can also  consider $so(3,1)$ real valued fields in order to
avoid reality conditions, and compare in the Lagrangian level with
the Holst action and in the Hamiltonian level with the Barbero
formulation in the pure bosonic case.\\
Also the introduction of matter into the action or the
generalization to more general internal gauge groups and
supergroups could be considered. The consequences of the presence
of a cosmological constant regarding the deformation of the
symmetry group of discretized models, such as consistent
discretization approach, as well as the consequences of
degenerated metrics.

%%%%%%%%%%%%%%%%%%%%%%%%%%%%%%%%%%%%%%%%%%%%%%%%%%%%%%%%%%%%%%%%%%%%%%%%%%%%%%%%%%%%%%%%%%%%%%%%%%%%%%%%%%%%%%%%%%%%%%%%%%%%%%%%%%%%%%%%%%%%%%%%%%%%%%%%
%%%%%%%%%%%%%%%%%%%%%%%%%%%%%%%%%%%%%%%%%%%%%%%%%%%%%%%%%%%%%%%%%%%%%%%%%%%%%%%%%%%%%%%%%%%%%%%%%%%%%%%%%%%%%%%%%%%%%%%%%%%%%%%%%%%%%%%%%%%%%%%%%%%%%%%%
%%%%%%%%%%%%%%%%%%%%%%%%%%%%%%%%%%%%%%%%%%%%%%%%%%%%%%%%%%%%%%%%%%%%%%%%%%%%%%%%%%%%%%%%%%%%%%%%%%%%%%%%%%%%%%%%%%%%%%%%%%%%%%%%%%%%%%%%%%%%%%%%%%%%%%%%
%%%%%%%%%%%%%%%%%%%%%%%%%%%%%%%%%%%%%%%%%%%%%%%%%%%%%%%%%%%%%%%%%%%%%%%%%%%%%%%%%%%%%%%%%%%%%%%%%%%%%%%%%%%%%%%%%%%%%%%%%%%%%%%%%%%%%%%%%%%%%%%%%%%%%%%%

\noindent \textbf{Acknowledgements}\\[1ex]
We thank O. Obreg\'on, C. Ramirez and M. Sabido for useful
discussions. The author acknowledges support from a CONACyT
scholarship (M\'exico) and PROMEP postdoctoral grant.

%%%%%%%%%%%%%%%%%%%%%%%%%%%%%%%%%%%%%%%%%%%%%%%%%%%%%%%%%%%%%%%%%%%%%%%%%%%%%%%%%%%%%%%%%%%%%%%%%%%%%%%%%%%%%%%%%%%%%%%%%%%%%%%%%%%%%%%%%%%%%%%%%%%%%%%%
%%%%%%%%%%%%%%%%%%%%%%%%%%%%%%%%%%%%%%%%%%%%%%%%%%%%%%%%%%%%%%%%%%%%%%%%%%%%%%%%%%%%%%%%%%%%%%%%%%%%%%%%%%%%%%%%%%%%%%%%%%%%%%%%%%%%%%%%%%%%%%%%%%%%%%%%
%%%%%%%%%%%%%%%%%%%%%%%%%%%%%%%%%%%%%%%%%%%%%%%%%%%%%%%%%%%%%%%%%%%%%%%%%%%%%%%%%%%%%%%%%%%%%%%%%%%%%%%%%%%%%%%%%%%%%%%%%%%%%%%%%%%%%%%%%%%%%%%%%%%%%%%%
%%%%%%%%%%%%%%%%%%%%%%%%%%%%%%%%%%%%%%%%%%%%%%%%%%%%%%%%%%%%%%%%%%%%%%%%%%%%%%%%%%%%%%%%%%%%%%%%%%%%%%%%%%%%%%%%%%%%%%%%%%%%%%%%%%%%%%%%%%%%%%%%%%%%%%%%

\end{document}